# Cyber Security in Cloud: Risk Assessment Models

Carlos Bendicho [1]

Independent ICT Researcher, ACM Member, IEEE Communications Society Member
FITCE Member, COIT Member, Spain
`carlos.bendicho@coit.es`

https://orcid.org/0000-0002-8538-0043

**Abstract.** The present paper shows a proposal of the characteristics Cloud Risk Assessment Models should have and presents the review of the literature considering those characteristics in order to identify current gaps. This work shows a ranking of Cloud RA models and their degree of compliance with the theoretical reference Cloud Risk Assessment model. The review of literature shows that RA approaches leveraging CSA (Cloud Security Alliance) STAR Registry that have into account organization´s security requirements present higher degree of compliance, but they still lack risk economic quantification. The myriad of conceptual models, methodologies and frameworks although based on current NIST SP 800:30, ISO 27001, ISO 27005, ISO 30001, ENISA standards could be enhanced by the use of techno-economic models like UTEM, created by the author, in order to conceive more simplified models for effective Risk Assessment and Mitigation closer to the theoretical reference model for Cloud Risk Assessment, available for all cloud models (IaaS, PaaS, SaaS) and easy to use for all stakeholders.

## 1 Introduction

Cloud Computing is nowadays a non-stoppable trend in marketplace that is impacting the shape and operation of all industries. The advent of SARS-Cov-2 pandemic has provoked an acceleration of cloud solutions adoption rhythm at a faster pace than all stakeholders foresaw.

Organizations movements towards cloud solutions in current context are not motivated only by original convictions but in some cases by urgent needs of coping with challenging changes in demand to scale out IT infrastructure fast and also scale it in when necessary, choosing subscription models based in OpEx (Operation Expenses) instead of CapEx (Capital Expenses) in order to grant Cash Flow and to be able to innovate fast and accelerate innovation cycles ("failing fast and cheap"). Most organizations with IT departments know and use virtualization in private clouds, but although many of them know containers technology, there is still a lack of expertise in using and operating containerized workloads as well as the different available choices in public cloud offerings such as Function Platform as a Service (FPaaS) or Code as a Service (CaaS) as well as the possibility of using Infrastructure as Code approaches.

---

[1] Dr. Carlos Bendicho holds M.Sc. and Ph.D. degrees in Telecommunications Engineering (EECS) from Bilbao School of Engineering, University of the Basque Country, Spain. He is also MBA from Technical University of Madrid and MIT Sloan Executive Program in Artificial Intelligence and Strategy, Massachusetts Institute of Technology.

Cloud adoption requires organizations to understand not only pure operational challenges but also to be aware of cyber security risks they face before and after migration and during cloud operation [22].

Assessment of Cloud adoption and migration for different workloads from a pure operational perspective is currently included in cloud ecosystem portfolio. This Cloud assessment can consider automatic IT infrastructure discovery and techno-economic evaluation of different cloud service providers identifying workloads, suitability, cost and effort of migration and operation, delivering a roadmap of different workloads migration. The assessment allows creating architectural foundations considering security and network perspectives, documenting them using Infrastructure as Code (e.g.: in JSON or YAML files) and leaving everything ready for deployment and workload migration.

However, as threat surface area is increased, organizations must conduct a complementary assessment of cyber security risks for all workloads before and after cloud migration with systematic reviews during operation in order to evaluate risks and adopt corresponding mitigation strategies [21].

As defined in [14, 26]: *"Cloud risk assessment is defined as a dynamic, step by step, repeatable process used to produce an understanding of cloud risks associated with relinquishing control of data or management of services to an external party (service provider)"*.

There are diverse cyber security risk assessment models and methodologies in the literature with different features and contributions to the state of the art which compound a very heterogeneous landscape that makes it difficult to standardize and industrialize the risk assessment process for cloud based solutions [14, 21]. There has also been a lack of systematic research on cloud security risk analysis and risk calculation as well as a trend towards more qualitative analysis or semi-quantitative, rather than quantitative research [23].

Therefore, there is a need for solvent cyber security risk assessment models that allow agile cyber risks assessments in cloud environments. These models should have into account the different market stakeholders' perspectives: infrastructure providers, service providers, enterprise end-customers, SME end-customers, government end-customers, considering all the supply chain actors [14]. They should be simple to implement and automatable for agile risk assessment. These models should also provide partial and overall risk evaluations in a quantitative way in order to prioritize risks and adopt adequate mitigation strategies.

This paper presents in Section 2 the review of relevant literature models along with conclusions of the review in section 2.2 and the statement of the research problem in section 2.3. Section 3 proposes the characteristics an agile cyber security risk assessment model should have. Section 4 shows a classification of the literature models according to the degree of compliance with the characteristics of the theoretical model, identifying current gaps. Section 5 discusses related work and Section 6 exposes conclusions and future work about using a universal techno-economic model created and developed by the author [24, 25] as a construct for a new simple, agile and Automatic Cloud Cyber Security Risk Assessment Model (ACCSRAM).

## 2 Review of the literature of Cloud Risk Assessment Models

Section 2.1 presents the review of relevant models of Cloud Risk Assessment in the literature. The search in the literature has been made using the keywords "Cloud Risk Assessment" from 2014 to 2020 in IEEE Xplore, Google Scholar, ScienceDirect, SpringerLink, Elsevier and ACM Direct Library databases selecting relevant papers that include Cloud RA models. There are references older than 2014 ([8][20]) that have been found from initial papers references. Section 2.2 shows the conclusions of the literature review and Section 2.3 states the research problem.

### 2.1 Models of the literature

Reference [1] is SPIRAL$^{SRA}$ Security Risk Assessment Framework based on Microsoft´s STRIDE threat model [2, 12] with six categories of threats: Spoofing(S), Tampering(T), Repudiation(R), Information Disclosure(I), Denial of Service(D), and Elevation of Privileges(E). It also considers Security Goals or Security Requirements of customer´s organization or the corresponding stakeholder whose perspective is being considered in the risk assessment. SPIRAL$^{SRA}$ framework includes 3 phases: (1) Elicitation, (2) Analysis and (3) Control as exposed in Fig.1.

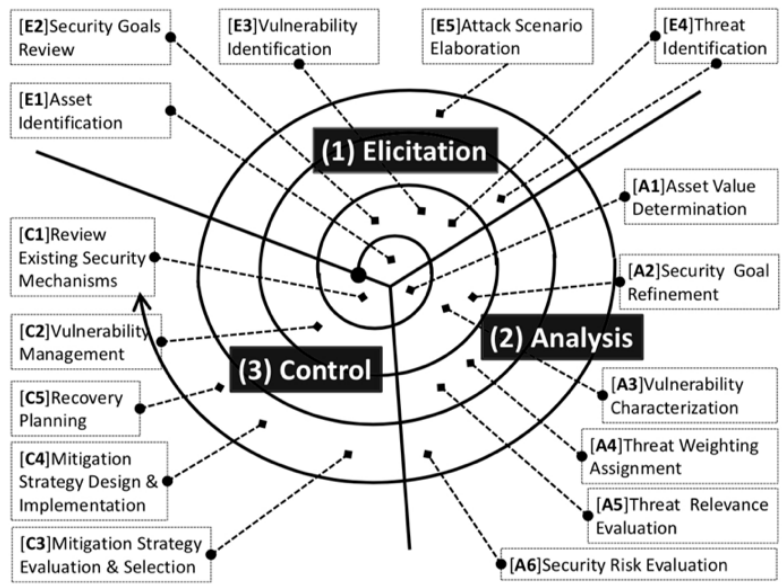

**Fig. 1.** Phases and Activities in the *SPIRAL$^{SRA}$* Security Risk Assessment Framework [1].

In [3], authors present a Risk Assessment Framework which provides Quantitative Risk Assessment (QRA) with a numerical estimate of the probability of a defined harm resulting from the occurrence of a particular event. This framework integrated in a software tool called OPTIMIS does not provide economic quantification of risks [4], [5]. It has into account Business Level Objectives (BLO), such as Quality of Service (QoS) or Service Level Agreement (SLA). It considers different end-users (service providers, brokers acting on their behalf, infrastructure providers and service consumers. OPTIMIS has into account the interaction between Service Provider and Infrastructure Provider including both perspectives, which allows to assess supply chain risks as well as different Cloud models (SaaS, PaaS, IaaS), multicloud, cloud federation, hybrid cloud, multicloud, private cloud, broker, cloud bursting. It provides quantitative risk assessment but no economic output.

Reference [6] is based on a risk matrix shown in Fig. 2. It quantifies partial and overall risk by assigning quantitative levels to assets considering incidence, threats as generation frequency and vulnerabilities in accordance to risk matrix before and after countermeasures, considering the following formula:

$$Risk\ (score) = Assets\ x\ Threats\ x\ Vulnerabilities \qquad (1)$$

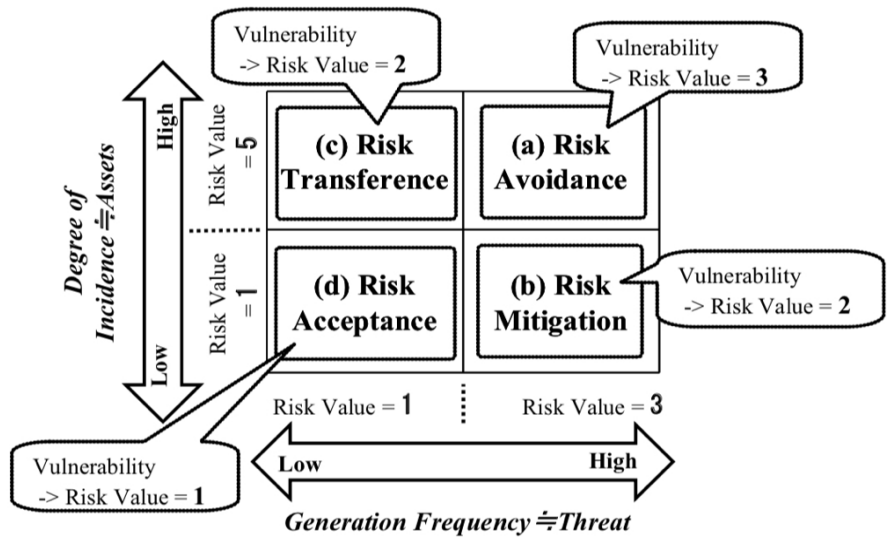

**Fig. 2.** Risk Matrix and value assignment to assets, threats and vulnerabilities [6].

In [7], authors propose Cloud Service Provider Risk Assessment Manager (CSPRAM) framework oriented for Cloud Service Providers (CSP) enabling Cloud Customers (CC) to contribute to the risk assessment process by defining risk factors in order to produce more realistic risk output. Fig. 3 depicts CSPRAM framework, which considers clients´security requirements for risk acceptance criteria and uses the following equation for risk score mapped on a risk matrix:

$$Risk\ (score) = Impact\ x\ Likelihood \qquad (2)$$

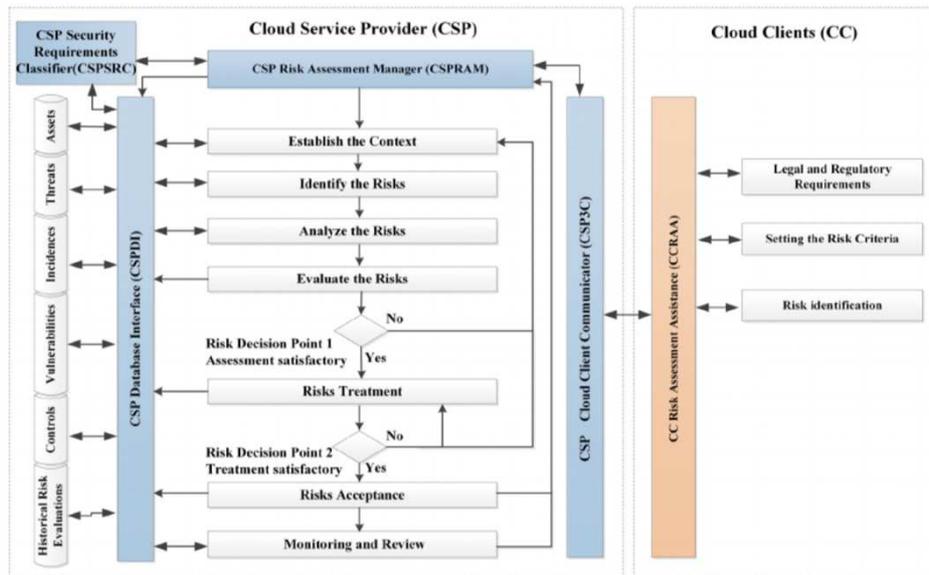

**Fig. 3.** CSPRAM Security Risk Assessment Framework [7].

Reference [8] presents a Quantitative Impact and Risk assessment framework for Cloud computing platforms called QUIRC. It uses the definition of risk as the product of the probability of a security threat event and its impact as in the following equation:

$$Risk\ (score) = Probability\ x\ Impact \qquad (3)$$

QUIRC bases probability assessment on security reports and uses wide-band Delphi method [9-11] of expert surveys to determine impact and weights for net risk calculation (overall risk) as a weighted average of risks. It also uses security objectives as those in Microsoft´s STRIDE threat model [2][12], the so-called CIAMAU framework [8] with six security objectives (First three from FISMA [13]: Confidentiality, Integrity, Availability, and last three proposed in [8]: Multi-party trust considerations, Mutual Auditability and Usability) or other security objectives (SO) models. An example of correspondence is shown in Table 1. The wide-band Delphi method allows the determination of weights for each of the six security objective categories.

**Table 1.** Correspondence between STRIDE and an example of other SO model [8].

| THREAT | EXAMPLE |
|---|---|
| **Spoofing:** adversary poses as a user or entity with an identity. [CONFIDENTIALITY] | Illegally using another user's authentication information, such as username and password. |
| **Tampering:** modification of data to achieve a malicious goal. [INTEGRITY] | Unauthorized changes to persistent data, or alteration of data over a network. |
| **Repudiation:** ability to deny a malicious action lacking proof. [AUDITABILITY] | User performs an illegal operation in a system that lacks the ability to trace it. |
| **Information Disclosure:** exposure of protected data to adversary. [CONFIDENTIALITY] | A cloud user reads a file from a co-tenant's workflow, without permission. |
| **Denial of Service:** adversary gains a higher trust level and attacks. [AVAILABILITY] | An adversary gains control of a tenant's VM, and makes another's Web server unavailable. |
| **Elevation of Privilege:** unprivileged user gains privileged access. [CONFIDENTIALITY] | An attacker penetrates all system defense to join the trusted system itself |

Reference [14] proposes the Cyber Supply Chain Cloud Risk Assessment (CSCCRA) model for CSPs identification, analysis and evaluation of cloud risks based on the dynamic supply chain. CSCCRA conducts an assessment of the cybersecurity posture of cloud suppliers prior to risk analysis phase. It uses a Multi-Criteria Decision-Making method (MCDM) to rank suppliers cybersecurity posture based on 52 security criteria grouped into 9 security target dimensions. A z-score is calculated for each target dimension and finally summed up for each supplier. Fig. 4 shows an overview of CSCCRA model. Risk output is a monetary value (cost) calculated with the following formula:

$$Risk(cost) = Impact(cost) \times Probability \times Frequency \quad (4)$$

Poisson distribution is used to represent the frequency of attack. Impact and Probability are represented by PERT distributions considering values of impact(cost) and probability provided by experts with and without mitigation controls. Risk value calculation is obtained by using Monte Carlo Simulation tool to combine stakeholders´ estimation of risk factors values and weighting.

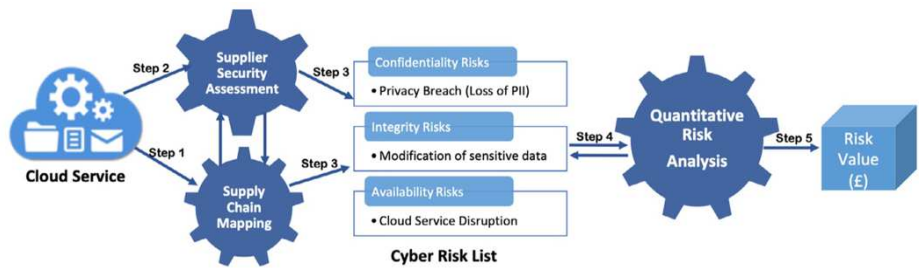

**Fig. 4.** Overview of CSCCRA model [14].

CARAM [15] is a qualitative risk assessment model based on previous ENISA [16], CAIQ and CNIL high-level frameworks. CARAM is oriented to provide Cloud Service

Customers (CSC) with a practical tool to select a Cloud Service Provider (CSP) that is the most compliant with customer´s risk profile. It allows a high degree of automation as it gets feed from CSA (Cloud Security Alliance) [17] Consensus Assessment Initiative Questionnaire (CAIQ) submitted by CSPs to CSA Star Registry available for CSCs [18] and based on CSA Cloud Control Matrix (CCM) taxonomy of security controls [19]. For the risk assessment, CARAM computes security, privacy and service risks for every CSC-CSP pair. CSCs need to specify acceptable risk levels for security, privacy and service categories, but do not need security expert level to specify those requirements.

Predecessor of CARAM is [20] that proposes a benchmarking framework for Cloud Security Level Agreements (Cloud SecLAs) using Quantitative Policy Trees (QPTs). It allows automation by gathering CSA Star Registry information from CSPs and benchmarking it vs Cloud Service Customer´s security requirements whose specification needs security expertise level. Based on this framework, [20] shows a Proof of Concept prototype of the framework called "QUANTifiable-Security-as-a-Service" system (QUANTSaaS). Fig. 5 depicts QUANTSaaS system building blocks.

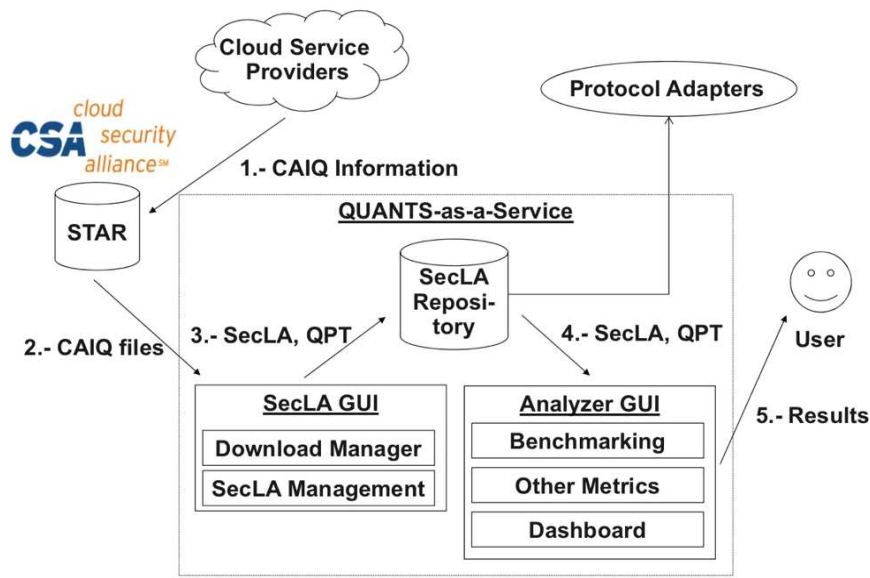

**Fig. 5.** Building blocks of the QUANTSaaS system [20].

Next Section 2.2 presents conclusions of the literature review which lead to the formulation of the research question in Section 2.3.

All the reviewed models are classified in Section 4 according to their degree of compliance with the characteristics of the theoretical reference model proposed in Section 3.

## 2.2 Conclusions of the literature review

Among the set of reviewed cyber risk assessment models in the literature, there is only one model that provides quantitative risk assessment with a monetary value for risk [14]. There are only two models that consider supply chain actors´security posture [3, 14]. Only a few consider Cloud Service Customer´s security requirements but need security expert level to specify them, so poor automation can be accomplished except in [20] which does not need security expert level.

In summary, there is a myriad of different cyber security risk assessment models with different features and contributions to the state of the art which compound a very heterogeneous landscape that makes it difficult to standardize and industrialize the risk asssesment process.

Hence, there is a need to select the most relevant features or characteristics a cyber risk assessment model should have in order to achieve an agile and effective risk assessment in cloud.

### 2.3 Research problem

Considering the conclusions of the literature review presented in the previous section, the research question is stated as follows:

*"Is it possible to design effective, standard, agile and automatic cyber security risk assessment models for cloud in order to industrialize risk economic evaluation in this domain?"*

In order to advance towards possible solutions for this research problem, next section presents the features or characteristics a theoretical Cloud Risk Assessment Reference Model should have.

## 3 Characteristics of the theoretical Cloud Risk Assessment Reference Model

The characteristics of a theoretical Cloud Risk Assessment Reference Model that allows agile risk assessment are proposed in this section as 12 characteristics named C1-C12. Acronyms are also included to make easy their identification in Table 2 (ST, SC, AC, EC, TH, SR, EA, SI, AU, BL, RM, CM).

The theoretical Agile Cloud Risk Assessment model should consider:

- C1: Different stakeholders`perspective: cloud service provider, infrastructure provider, customer organization, end-user, broker, government, regulation authority (Acronym ST in Table 2).
- C2: Supply chain actors and different components perspective (Acronym SC).
- C3: All cloud models (IaaS, PaaS, SaaS), cloud federation, multicloud, hybrid cloud, private cloud, broker, cloud bursting (Acronym AC).
- C4: Economic risk quantification (Partial risks and overall) (Acronym EC).
- C5: Threat based / Relevant Threats / Threat Prioritisation (Acronym TH).
- C6: Security Requirements or Security Objectives (Customer´s) (Acronym SR).
- C7: Elicitation / Analysis / Control / Mitigation / Review (Acronym EA).
- C8: Simplicity / To be as simple as possible (Acronym SI).
- C9: Automatable / Automatic assessment (Acronym AU).
- C10: Business Level Objectives (BLO) (Acronym BL).
- C11: Risk Matrix (Assets, Vulnerabilities, Threats) (Acronym RM).
- C12: Countermeasures for risk mitigation (Acronym CM).

Section 4 shows the classification of relevant papers in the literature according to their degree of compliance with these 12 characteristics proposed for the theoretical model, identifying current gaps for future work.

# 4   Classification of Cloud Risk Assessment Models

This section shows the classification of the literature models reviewed in previous section 2.1 according to their degree of compliance with the characteristics proposed in section 3 for the theoretical Cloud Cyber Security Risk Assessment Reference model.

Table 2 shows characteristics C1-C12 (in columns) with their acronyms for the Theoretical Model and 8 relevant models of the literature (in rows). Table 2 assigns 1 value to a cell if the model in the row fulfills the characteristic of that column and no value (0 value) if the model is not compliant with that characteristic. Total compliance for each model is summed up in last column (Total).

**Table 2.** Classification of models in the literature according to the characteristics established for the theoretical model.

| Characteristics | C1 ST | C2 SC | C3 AC | C4 EC | C5 TH | C6 SR | C7 EA | C8 SI | C9 AU | C10 BL | C11 RM | C12 CM | Total |
|---|---|---|---|---|---|---|---|---|---|---|---|---|---|
| Theoretical Model | 1 | 1 | 1 | 1 | 1 | 1 | 1 | 1 | 1 | 1 | 1 | 1 | 12 |
| Cayirci et al., 2016 CARAM [15] | 1 |  | 1 |  |  | 1 |  | 1 |  | 1 | 1 | 1 | 7 |
| Luna et al., 2012 QUANTSaaS [20] | 1 |  | 1 |  |  | 1 |  | 1 |  | 1 | 1 | 1 | 7 |
| Saripalli & Walters, 2010 QUIRC [8] | 1 |  | 1 |  | 1 | 1 |  |  |  | 1 | 1 | 1 | 7 |
| Akinrolabu et al., 2019 CSCCRA [14] | 1 | 1 | 1 | 1 |  |  |  |  |  |  |  | 1 | 5 |
| Nhalabatsi et al., 2018 [1] | 1 |  |  |  | 1 | 1 | 1 |  |  |  |  | 1 | 5 |
| Djemame et al., 2016 OPTIMIS [3] | 1 | 1 | 1 |  |  |  |  |  |  | 1 |  | 1 | 5 |
| Tanimoto et al., 2014 [6] |  |  |  |  | 1 |  |  | 1 |  |  | 1 | 1 | 4 |
| Albakri et al., 2014 CSPRAM [7] | 1 |  |  |  |  | 1 |  |  |  |  | 1 | 1 | 4 |

Table 2 shows that CARAM [15], QUANTSaaS [20] and QUIRC [8] models are the most compliant models from those analysed in the literature with 7 points out of 12 for the theoretical reference model what means a 58,33% of compliance. This leaves a gap of 41,66% for future models.

Both CARAM [15] and QUANTSaaS [20] leverage CSP submitted information to CSA STAR Registry, which allows high degree of automation but they still lack risk economic quantification. QUIRC [8] is a threat-based model that uses security objectives and Multi-Criteria Decision-Making method (MCDM) for risk economic quantification providing monetary value for risk, but more difficult to automate.

Thus, the characteristics to improve and bridge the mentioned compliance gap of 41,66% for future models considering these TOP 3 models are C2 (SC: Supply Chain Actors and different components perspective), C4 (EC: Economic risk quantification (partial and overall) including monetary measures, C5 (TH: Threat based / Relevant Threats / Threat Prioritisation), C7 (EA: Elicitation / Analysis / Control / Mitigation / Review) having into account the whole cycle periodical review, C8 (SI: Simplicity) and C9 (AU: Automatable / Automatic Assessment) in order to make true an agile cyber risk assessment in cloud.

## 5 Related Work

A few papers in the literature conduct a classification of cloud risk assessment models and methodologies. In [21], authors classify cloud risk assessment methods and tools considering five categories: risk assessment as a service (RAaaS), quantitative or qualitative analysis, graphs analysis, hierarchical assessment and security matrix assessment, describing the type of risk modeling, identifying contemplated stakeholders and concluding that basically studied RA methods use qualitative and/or quantitative approach. Reference [22] adds another category: semi-quantitative analysis from [27, 28] and compares several methods contemplating stakeholders, automation, efficiency, adaptability, collaboration and support tool. In [14], authors present a comparison of cloud RA models emphasizing adopted assessment method (quantitative, qualitative or semi-quantitative), use of external experts and supply chain associated risks inclusion.

This paper proposes the features of a theoretical agile cyber security risk assessment reference model based on the literature review and author´s professional experience in ICT and cloud computing industry. The subsequent classification of relevant papers of the literature according to the degree of compliance with those features allows to identify current gaps for further improvement in future models.

## 6 Conclusions and Future Work

This paper has proposed the characteristics a theoretical agile Cloud Cyber Security Risk Assessment reference model should have. After reviewing relevant papers of the literature, they have been classified based on their degree of compliance with the characteristics of the theoretical reference model. The three most compliant models of the literature present 58,33% of compliance, leaving a gap of 41.66% for future models to accomplish further improvement. From the analysis of the gap in characteristics, this improvement should be focused on economic risk quantification, supply chain actors and components perspective, threat prioritisation, whole cycle periodical review, simplicity and automatic assessment in order to achieve agile risk assessment in cloud cyber security.

Author´s future work in this direction will consider to use a universal techno-economic model called UTEM created and developed by the author [24, 25] as a construct for a simple, agile and Automatic Cloud Cyber Security Risk Assessment Model (ACCSRAM).

UTEM model is a universal, flexible, scalable and generalizable model that allows to industrialize techno-economic assessment in multiple domains considering all market players perspectives, also suitable for technological consulting and currently available for all industry stakeholders under specific license of use [24, 25].